\documentclass{aa}  

\usepackage{txfonts}
\usepackage{graphicx}	
\usepackage{amsmath}	
\usepackage{multirow}
\usepackage{subfigure}
\usepackage{adjustbox}
\usepackage{xcolor}
\usepackage{float}
\usepackage{amsmath}	
\usepackage{lscape} 
\usepackage{caption}
\usepackage[switch]{lineno}
\usepackage{natbib}
\usepackage[colorlinks=true,linkcolor=blue,allcolors=blue]{hyperref}
\newcommand{\fl}{\emph{Fermi}-LAT~}

\newcommand{\gam}{$\gamma$}

\usepackage{multirow}

\begin{document}

\title{Search for $\gamma$-ray emission from SNRs in the Large Magellanic Cloud: a new cluster analysis at energies above 4 GeV}

   \subtitle{}
\author
 {A. Tramacere\inst{1} $\thanks{E-mail: \url{andrea.tramacere@unige.ch}}$
  \and
  R. Campana \inst{2}
  \and
  E. Massaro \inst{3}
  \and
  F. Bocchino\inst{4}
  \and
  M. Miceli \inst{4,5}
  \and
  S. Orlando \inst{4}}
\institute{
 Department of Astronomy, University of Geneva, Chemin Pegasi 51, 1290, Versoix, Switzerland
 \and
 INAF/OAS, via Gobetti 101, 40129, Bologna, Italy
 \and
 INAF/IAPS, via del Fosso del Cavaliere 100, I-00113 Roma, Italy 
 \and
 INAF/Osservatorio Astronomico di Palermo, Piazza del Parlamento 1, I-90134 Palermo, Italy
 \and
 Dipartimento di Fisica e Chimica E. Segr\`e, Universit\`a di Palermo, Piazza del Parlamento 1, 90134, Palermo, Italy
}
   \date{Received ...; accepted ...}
\label{firstpage}


   \abstract
   {A search for \gam-ray emission from SNRs in the Large Magellanic Cloud (LMC) based on the detection of concentrations in the arrival direction Fermi-LAT images of photons at energies higher than 10 GeV found significant evidence for 9 of these sources.
This analysis was based on data collected in the time window since August, 4 2008 to August, 4 2020 (12 years)
   }
   {In the present contribution, we report  results of a new search extended using a 15-year long (up to August, 4 2023) data set and to a broad energy range (higher than 4 GeV). 
   The longer baseline and the softer energy lower limit are required to further understand the relation between the X-ray and gamma ray SNRs in the LMC, and to investigate the completness of the sample at low luminosities.
    }
   {Two different methods of clustering analysis were applied: Minimum Spanning Tree (MST), and the combination of Density-Based Spatial Clustering of Applications with Noise (DBSCAN) and DENsity-based CLUstEring (DENCLUE) algorithms. 
   }
   {We confirm all previous detections and found positive indications for at least 8 new clusters with a spatial correspondence with other SNRs, increasing thus the number of remnants in LMC candidate or detected in the high energy \gam-rays to 16 sources.
   }
    {This study extends previous analyses of $\gamma$-ray emission from SNRs in the LMC by incorporating a longer observational baseline and a broader energy range. The improved dataset and advanced clustering techniques enhance our understanding of the connection between X-ray and $\gamma$-ray SNRs, providing new insights into their high-energy properties, and contributing to assessing the completeness of the sample at lower luminosities.}

   \authorrunning{A. Tramacere et al.}
\titlerunning{SNRs in the Large Magellanic Cloud at energies above 4 GeV}

\maketitle


\section{Introduction}

The Large Magellanic Cloud (LMC) is a nearby irregular galaxy particularly rich of Supernova Remnants (SNRs), with ages in the interval from about 40 years to a few tens of thousands of years. 
The LMC provides, therefore, a unique laboratory to investigate high energy processes related to the interaction of shock waves with the interstellar gas and particle acceleration.
 In particular, these energetic processes can be traced by the detection of \gam-ray emission from GeV to TeV energies.

The first observation of the LMC at energies higher than 100 MeV was done by \citet{sreekumar92}, on the basis of \emph{Compton Gamma Ray Observatory} (CGRO) EGRET data. 
Later, \citet{abdo10} reported a first study of the LMC using the 11-month \emph{Fermi}-Large Area Telescope (LAT, \citealt{ackermann12}) data and discovered a bright extended high energy source associated with the massive star forming region 30 Doradus. 
\citet{ackermann16} performed the first detailed analysis of the LMC region considering a much richer \fl data spanning the first 6 years of the mission and detected four distinct \gam-ray sources. Since then, only a few research papers have been published about the discovery of SNRs in the \gam-ray band.
The last version (DR4, v34) of the 4FGL catalog released by the \fl collaboration \citep[]{abdollahi20,ballet20} contains 18 sources, detected in the 50 MeV -- 1 TeV energy range. Of these, 4 are reported as extended features and 7 are present also in the 3FHL catalog \citep{ajello17} at energies above 10 GeV.
Furthermore, possible background extragalactic counterparts (unclassified blazar or BL Lac objects) are indicated for 9 of these 4FGL sources.

At TeV energies, the  H.E.S.S. collaboration observed three sources up to about 10~TeV in the LMC \citep{abramowski15}, two of them associated with the SNRs N~157B \citep{abramowski12} and N~132D, the former containing the highest known spin-down luminosity pulsar PSR J0537$-$6910 \citep{marshall98,cusumano98} with the fast period 
of 16 ms.

\citet{campana18}, hereafter Paper I, used the data collected by the \fl telescope in 9 years of operation for a photon cluster search at energies higher than 10 GeV, and reported the discovery of high energy emission from the two SNRs N~49B and N~63A.
In a subsequent work (\citealt{campana22}, Paper II) a more complete search was applied to the Fermi-LAT data over 12 years with energy higher than 6 and 10 GeV and using two cluster-finding methods.
Detections of paper I were confirmed and new significant clusters associated with the SNRs N49 and N44 were found, furthermore, another cluster likely associated with N186D was detected at energies higher than 6 GeV. The first two sources are among the 
brightest X-ray remnants in the LMC and correspond to core-collapse supernovae 
interacting with dense H II regions, indicating that a hadronic origin of high-energy 
photons are the favourite process. Furthermore, the discovery above 6 GeV of a cluster 
corresponding to the not-very-bright SNR N186D suggested that other remnants of comparable luminosity can be found extending the search at lower energies, thus calling for an extension of the Paper I-II analysis with a longer baseline and a softer energy lower limit

In the present paper we extend the analysis of the LMC region using a richer data set of 15 years of \fl observation and energies higher than 4 GeV.
As in Paper II we apply the two cluster-finding methods DBSCAN/DENCLUE \citep{Tramacere:2013,Tramacere:2016} and Minimum Spanning Tree (hereafter MST, see \citealt{campana08,campana13} for recent use in \gam-ray astronomy) for extracting photon spatial concentrations likely corresponding to genuine \gam-ray sources. The latter method has been already successfully applied in several works \citep{bernieri13,paperI,paperII,paperIII,paperIV,campana17,
campana18b,campana21}.

The results of this new analysis confirm the previous SNR detections with a higher significance. Moreover, some other photon clusters are found at positions very close to those of SNRs not previously associated with \gam-ray emitters.

The outline of this paper is as follows. 
In Section~\ref{s:dataLMC} the data selection and the MST and DBSCAN/DENCLUE algorithms are described, while the analysis of the LMC region is presented in Section~\ref{s:alg}. 
The present status of knowledge on the high energy sources in LMC, mainly based on the Fermi catalog, is summarized in Section~\ref{s:clF4} and in Section~\ref{s:cl410} we deepen the analysis of some interesting SNR counterparts to photon clusters, while in Section~\ref{s:conc} the results are summarized and discussed.

\section{Data selection in the LMC region}     
\label{s:dataLMC}

We analyzed 15 years of \fl data collected from August 4, 2008, to August 4, 2023, including events above 4 GeV.
The complete dataset was retrieved as weekly files from the FSSC archive\footnote{\url{http://fermi.gsfc.nasa.gov/ssc/data/access/}}. The dataset was processed using the Pass 8, release 3, reconstruction algorithm and includes the instrumental responses.
The event lists were then filtered applying the standard selection criteria on data quality and zenith angle (source class events, \texttt{evclass} 128), front and back converting (\texttt{evtype} 3), up to a maximum zenith angle of $90\degr$. 
Events were then screened for standard good time interval selection.
 We selected a region $12\degr\times9\degr$, in Galactic coordinates $271\degr < l < 287\degr$ and $-38\degr < b < -29\degr$ and approximately centered at the LMC, and the two cluster-finding algorithms were applied. As in Paper I, the size of this region was chosen to include all the LMC and its near surroundings, but here the extension in longitude is wider  than the previous one to enhance the cluster extraction at low energy because of the higher mean angular distance between photons 

Given the condition of a high local diffuse emission in the LMC, the sorting of point-like sources is a quite delicate work, and particular approaches should be adopted. 
Our basic choice is to search clusters at energies higher than 10 GeV, since the diffuse component is rather weak in this band.
Once a list of clusters is obtained, we searched for correspondent features at lower energies and, if they were found, the cluster was considered to be confirmed.
However, in these analyses we had to use smaller fields not including the 30 Dor nebula, because its relatively high number of photons produces a too short mean angular separation in the field for a safe cluster search.
Analyses at lower energies generally imply that the resulting \emph{cluster parameters} (as defined in the following) are generally much lower than above 6 GeV, and the extraction of clusters is less stable: in particular, small changes in the selection parameters could give different structures.

The LMC is particularly rich of SNRs.
The work of \citet{maggi16} reports 59 remnants.
A richer catalog, which includes the previous one, adding another group of 15 SNR candidates, is provided by \citet{bozzetto17}; a further sample of 19 optically selected SNR candidates is given by \citet{yew21}.
A new complete catalog has been recently published by \citet{eros24} based on eROSITA X-ray survey and including 78 entries (hereafter ZMC24).
As it will be clear from the results of this work, we did not find any well established positional correspondence between clusters and the \citet{yew21} sources, the \citet{bozzetto17} candidate sources, and new additions of eROSITA to the list of \citet{maggi16}, therefore we will limit our analysis to the last sample, which contains the brightest and genuine SNRs.

\begin{table*}
\caption{SNRs already detected at energies higher than 4 and 10 GeV and confirmed by the present analysis.}
\begin{center}
\begin{tabular}{lcccrrrrrrl}
\hline
  SNR        & RANGE   &  TYPE  &  D  &   \multicolumn{1}{c}{$L_X$}    & $S_{\rm cls }4$      & $S_{\rm cls }10$      & $\Delta$  & MST $N/M$ & MST $N/M$ & Notes \\
             & ~ GeV   &        & $'$ &   \multicolumn{1}{c}{$10^{35}$ erg s$^{-1}$}       &         &         &   $'$     &   4 GeV ~~& 10 GeV  ~~&        \\  
\hline
   &  \\

  N~44       & $>$4;10  &  --    & 4.3 &  0.90   & 4.8   & 3.3   &  3.7  & 26/48.1 & 13/28.7 &        \\ 
  N~132D     & $>$4;10  & CC     & 2.1 & 315.04  & 6.6   & 4.6   &  1.8  & 27/55.6 & 21/57.6 &   4FGL \\ 
  N~49B      & $>$4*;10 & CC     & 2.8 & 38.03   & *5.6  & *3.9  &  3.4  & *23/46.3 &  6/20.0 &   Nf     \\ 
  N~49       & $>$4*;10 & CC-SGR & 1.4 & 64.37   & *5.6  & *3.9  &  3.5  & *23/46.3 &  9/18.4 &   Nf      \\ 
  B0528$-$692 & $>$10  &         & 4.5 &  1.99   & 4.4   & 3.5   &  4.1  & 24/55.7 & 24/58.8 &        \\ 
  N~63A      & $>$4;10  & CC     & 1.4 & 185.68  & 6.5   & 4.1   &  0.4  & 33/92.9 & 15/53.7 &   4FGL \\ 
  N~157B     & $>$4;10  & CC-PWN & 2.0 & 15.00   & 18.8  & 11.6  &  1.6  & 146/441.5  & 155/577.4 & Nf30D~ 4FGL \\ 
  B0540$-$693 & $>$10  & CC-PSR  & 1.2 & 87.35   & 8.7   & 4.8   &  5.3  & 22/61.8 &  8/35.4 & Nf30D~ 4FGL \\ 
             &          &        &     &         &       &       &       &   &     &        \\  
  N~186D     & $>$4;6    &        & 1.9 &  1.09   & ---   & 3.7  &       & 9/17.1 &  7/19.0 &        \\  
\hline
\end{tabular}
\tablefoot{
* : unresolved cluster at energies lower than 10 GeV \\
D : SNR diameter \\
$\Delta$ : angular separation \\
$L_X$ : X-ray luminosity in the band 0.3–8 keV from the Maggi et al. (2016) catalogue\\
$S_{\rm cls}$: significance of the DBSCAN clusters at energies higher than 4 and 10 GeV.  \\
$N/M$: number of photons in the clusters and the corresponding magnitudes from MST search.
}
\end{center}

\end{table*}

\section{Description of photon cluster detection algorithms}    
\label{s:alg}

\begin{table*}
\begin{center}

\caption{New SNRs detected at energies higher than 4 GeV.}
\begin{tabular}{lccrrrlrll}
\hline
  SNR        & $l$ & $b$  &  TYPE  &  D  & {$L_X$}  & $S_{\rm cls }4$  & $\Delta$ &  MST &  Notes \\
             & deg & deg  &        & $'$ &          &                  &  $'$     & $N/M$ &        \\  

\hline
   &  \\
  B0453$-$685 & 279.793 & $-$35.765  &  CC    & 2.0 & 13.85   & 3.7 & 2.0  &  10/18.3 &  \\  
  B0532$-$675 & 277.638 & $-$32.346  &        & 4.7 &  2.48   & 3.1 & 5.8  &  7/23.5 &  e \\  
  \lbrack HP99\rbrack 1139  & 281.589 & $-$34.229  &        & 4.4 &  1.44   & 3.1 & 6.4  &  13/24.7 & e \\  
  N~103B      & 279.624 & $-$34.401  &  Ia    & 0.5 & 51.70   & 3.3 & 2.7  &  10/19.8 &  \\      
  B0519$-$690 & 279.617 & $-$33.323  &  Ia    & 0.6 & 34.94   & 4.2 & 5.0  &  13/26.6 &  \\  
  DEM~L316A/B & 279.984 & $-$30.859  &  Ia/CC & 3.2 &  1.47/1.26   & 4.7 & 4.1  &  11/21.5 & Nf \\  
\hline
   &  \\
  DEM~L241    & 277.712 & $-$32.150  &        & 5.3 &  3.84   & 5.2 & 1.2  &  35/79.6 &  \\  
\hline
\end{tabular} \\
\tablefoot{
Galactic coordinates and angular separations are from MST, the X-ray luminosity is in units of 10$^{35}$ erg/s.\\
e : MST analysis at energies $>$6 GeV.\\
Nf : M value obtained in a field with $b > -31^o$.\\
$S_{\rm cls}$: significance of the DBSCAN clusters at energies higher than 4 GeV.  \\
$N/M$: number of photons in the clusters and the corresponding magnitudes from MST search.
}
\end{center}
\end{table*}

\subsection{DBSCAN and DENCLUE}
\label{ss:dbs}
DBSCAN is a density-based clustering method that uses the local density of points to find clusters in data sets that are affected by background noise.
Here we give only a brief description of the method and its application,
for a detailed description see \cite{Tramacere:2013}.
Let $D$ be a set of photons, where each element is described by the sky coordinates. 
The distance between two given elements $({p_l,p_k})$ is defined as the angular 
distance on the unit sphere, $\rho(p_l,p_k)$. 
Let $N_{\varepsilon}(p_i)$ be the  set of points contained within a radius $\varepsilon$, centred on  $p_i$, and $|N_{\varepsilon}(p_i)|$ the number of contained points, 
i.e., the estimator of the local density, and $K$ is the threshold value such that $|N_{\varepsilon}(p_i)|\geq K$ defines \emph{core} points.
The DBSCAN builds the cluster by recursively connecting \emph{ density connected}  points to each set of \emph{core} points found in $D$. 
Each cluster $C_m$ will be described by the position of the centroid ($x_c$,$y_c$), the ellipse of the cluster containment, the number of photons in the cluster ($N_p$), and  the significance.
The ellipse of the cluster containment is evaluated using the principal component analysis method.
The significance of a cluster ($S_\mathrm{cls}$), evaluated  according to the 
signal-to-noise  Likelihood Ratio Test (LRT) method proposed by \cite{Li&Ma},  
and explained in detail in \cite{Tramacere:2013} (see also Paper II). 
To optimize the clustering process, we need to set proper values for both  $\varepsilon$ and $K$, we proceed as follows:
\begin{enumerate}
\item we set the radius $\varepsilon$ to be half of the instrument point spread function (PSF) at a given threshold energy
\item we evaluate the initial value of the background density level $\eta_{\rm bkg}$ equal to the number of photons within the field divided by the field angular extent, and we set $K=N_{\rm rej}\eta_{\rm bkg}\Omega_{\varepsilon}$ where $\Omega_{\varepsilon}$ is is the solid angle encompassed by the radius $\varepsilon$, and $N_{\rm rej}$ is the background rejection factor, that we set to 4, to get a conservative rejection level at 2-$\sigma$ confidence level.
\item we perform a first DBSCAN detection using the parameters obtained at step 2. We update the value $\eta_{\rm bkg}$ by removing the angular extent of each detected cluster from the field angular extent, and the number of corresponding photons, obtaining a new value  $\eta_{\rm bkg}^*$
\item we perform a second DBSCAN detection updating $K$ with the updated $\eta_{\rm bkg}^*$.
\item we evaluate the significance of each cluster $S_\mathrm{cls}$ defined as
\begin{equation}
       S_{\rm cls}=\sqrt{2 \left  ( N^{\rm in}_{\rm src} \ln \left[ \frac{2 N^{\rm in}_{\rm src}}{N^{\rm in}_{\rm src}+N^{\rm eff}_{\rm bkg}} \right]+
       N^{\rm eff}_{\rm bkg} \ln \left[ \frac{2N^{\rm in}_{\rm src}}{ N^{\rm in}_{\rm src}+N^{\rm eff}_{\rm bkg}} \right ]  \right  ).}
       \label{eq:signif}
\end{equation} 
where $N^\mathrm{in}_\mathrm{src}$ is the number of points in each cluster, and $N^\mathrm{eff}_\mathrm{bkg}$ is the expected number of background photons within the circle enclosing the most distant cluster point, according to  value of $N_\mathrm{bkg}$ estimated from $\eta_{\rm bkg}^*$, and then we remove those below the threshold value
\end{enumerate}

When sources are very close, the DBSCAN algorithm might not be able to separate them. The possibility of blended sources increases when the density of background photons exhibits gradients requiring different values of $K$ across the same sky region. This was not the case in the analysis presented in Paper I, where we used a threshold of 6 GeV, but became more relevant in the current analysis by reducing the threshold to 4 GeV.
To deblend two (or more) `confused' sources we use the \emph{DENsity-based 
 CLUstEring} (DENCLUE) algorithm \citep{Hinneburg:1998wo,Hinneburg:2007tq}.   

  DENCLUE relies on kernel density estimation, defining clusters according to the local maxima of an estimated density function. 
 This function, $f(p)$, based on a Gaussian kernel in the current application,  represents the sums of the  \emph{influence}  of each in a data set $\{p_1,...,p_j\}$, on the position  $p$ of the data space. a sub-set of points converging to the same local maximum defines a cluster. 
 This is accomplished by finding, for each point $p_j$ in the dataset, the corresponding   {density attractor}  point $p_j^*$, i.e. a local maximum of $f(p)$. 
 Hence, by clustering the {density attractors}, each cluster of {density attractors}, $A_{C_n}$, will identify a common convergence point for the corresponding points $\{p_k:p^*_k \in A_{C_n}\}$, i.e. $\{p^*_k \in A_{C_n}\}$ will map a new sub-cluster of points $\{p_k \in C_{D_n}\}$.
An application of the DENCLUE algorithm for deblending confused sources is given in \cite{Tramacere:2016}, and a detailed description of this algorithm is reported in Appendix \ref{denclue-appendix}.
The deblending process, applied to each cluster detected by the DBSCAN, proceeds as follows:
\begin{enumerate}
\item we set the values of the kernel width $h$ according to the instrument point spread function (PSF) at a given threshold energy.
\item each cluster $C_{\rm m}$, detected by the DBSCAN, defines a set of points $\{p_j \in C_{\rm m}\}$, corresponding to the coordinates of the photons in the cluster and is processed by the DENCLUE
\item all the attractors points $p_j^*$  found in $C_{\rm m}$ are clustered using the DBSCAN, providing a list of clusters of \emph{density attractors} $\{A_{C_1},...,A_{C_n}\}$ and mapping a corresponding list of deblended sub-clusters $\{C_{D_1},...,C_{D_n}\}$. Of course, if a DBSCAN cluster, $C_m$, will result in a single cluster of attractors, $\{A_{C_1}\}$, then $C_m$  will not be partitioned in sub-clusters, i.e. $C_{D_n} \equiv \{A_{C_1}\}\equiv C_m$
\item Each sub-cluster $C_{D_n}$ is eventually validated or discarded according to its cluster significance evaluated as in Eq \ref{eq:signif}.
\end{enumerate}

\begin{figure}
   \centering
   \resizebox{\hsize}{!}{
   \includegraphics[angle=0]{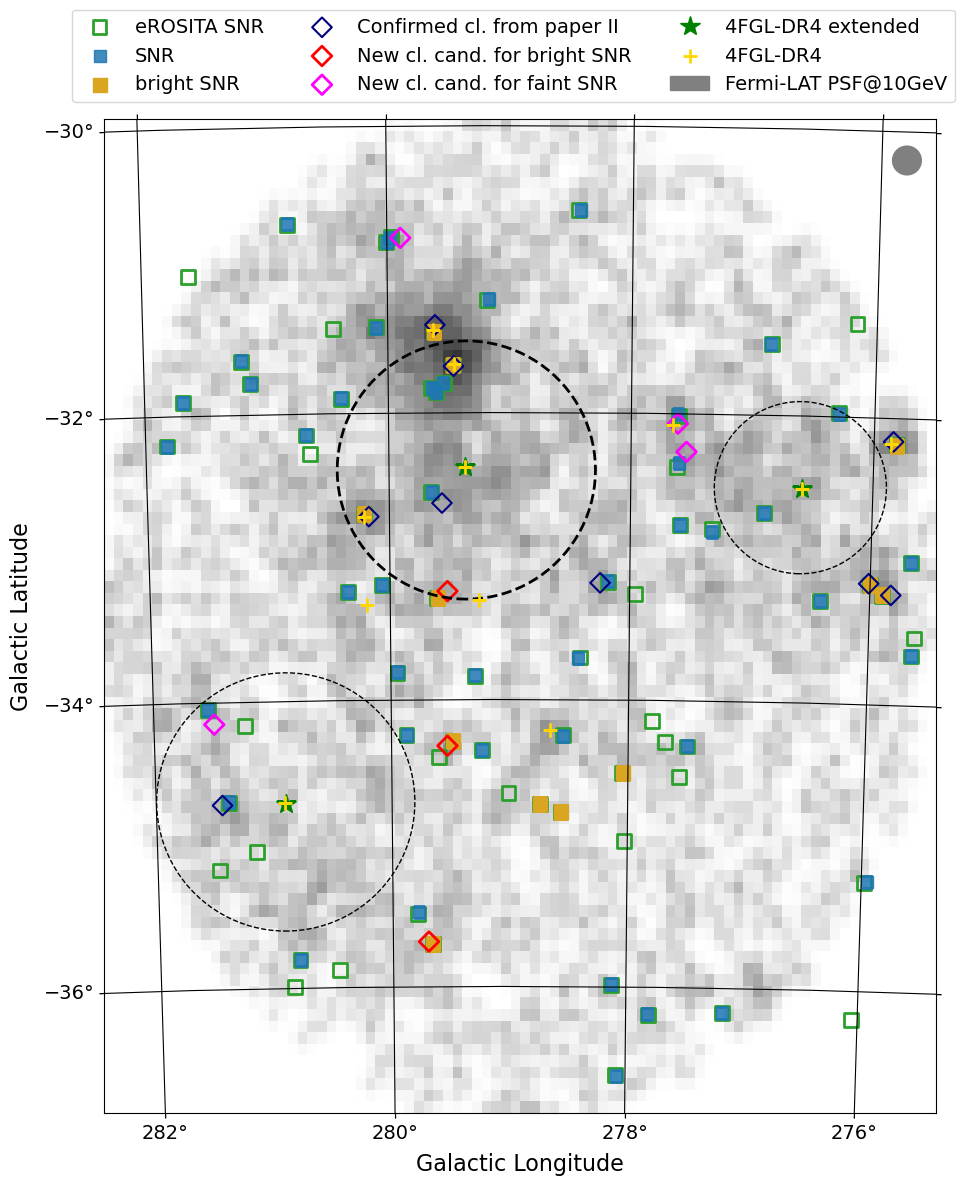}}
      
      \caption{Map of the region of the Large Magellanic Cloud reporting the known SNRs, 4FGL-DR4 sources, and the photon clusters found in the present work. Solid squares identify SNRs from \cite{maggi16}, green empty squares SNRs from \cite{eros24}.
      Blue diamonds identify confirmed clusters form \cite{paperII}, red and purple diamonds identify clusters detected in the present analysis for bright and faint SNRs, respectively.
      Yellow crosses mark 4FGL-DR4 point-like sources,  dashed circles represent the approximate size of the extended 4FGL sources, the thick one is that containing the 30 Dor complex, and green stars mark their centres. 
      }
    \label{Fig1}
\end{figure}

\subsection{Minimum Spanning Tree}
\label{ss:mst}

The MST method is a topometric tool useful for searching spatial concentrations in a field of points (see, for instance, \citealt{cormen09} and also \citealt{campana08,campana13}).
The application of this method to the $\gamma$-ray sky and selection criteria were presented elsewhere (e.g., in \citealt{campana18}). 
We, therefore, summarise here only how the MST parameters were optimized to enhance the detection of photon clusters above 4 GeV in the LMC region.

Consider a two-dimensional set of $N_n$ points (\emph{nodes}), there is a the unique \emph{tree} (a graph without closed loops) connecting all the nodes with the minimum total weight, $\min [\Sigma_i \lambda_i]$, where the elements of the set $\{\lambda\}$ are, in this case, the angular distances between photon arrival directions. 

The \emph{primary} cluster selection is then performed in two following steps: the elimination of all the edges having a length $\lambda \geq \Lambda_\mathrm{cut}$, that it is convenient to give in units of the mean edge length $\Lambda_m = (\Sigma_i \lambda_i)/N_n$ in the MST, and the removal of all the sub-trees having a number of nodes $N \leq N_\mathrm{cut}$. This leaves only the clusters with a number of photons higher than this threshold. 
For each cluster we evaluated the following parameters: \emph{i}) the number of nodes in the cluster $k$, $N_k$; \emph{ii}) the \emph{clustering parameter} $g_k$, defined as the ratio between $\Lambda_m$ and $\lambda_{m,k}$, the mean length over the $k$-th cluster edges; \emph{iii}) the centroid coordinates, computed by means of a weighted mean of the cluster photon arrival directions \citep[see][]{campana13}; \emph{iv}) the radius of the circle centred at the centroid and containing the 50\% of photons in the cluster, the \emph{median radius} $R_m$ (for a genuine point-like $\gamma$-ray source should be smaller than or comparable to the 68\% containment radius of instrumental Point Spread Function or PSF); \emph{v}) the \emph{maximum radius} $R_\mathrm{max}$ as the distance between the centroid and the farthest photon in the cluster (usually of some arcminutes for a point-like source and a few tens of arcminutes either for extended structures or for unresolved close pairs of sources).
A very useful parameter here is the cluster \emph{magnitude}:
\begin{equation}
M_k = N_k g_k  ~~~~~~~~~.
\end{equation}

\cite{campana13} found that $\sqrt{M}$ is well correlated with other statistical source significance parameters, in particular with Test Significance from the Maximum Likelihood \citep{mattox96} analysis. 
The use of these parameters depends on the structure of the sky region under consideration: for example, in fields with a non-high spatial photon density, like those at energies above 10 GeV, a lower threshold value of $M$ around 15--20 would reject the large majority of spurious low-significance clusters, while in dense fields $M$ values lower than 15 can correspond to significant clusters.

\section{LMC sources in 4FGL-DR4 catalogue}   
\label{s:clF4}

There are 13 sources in the 4FGL-DR4 catalog of the Fermi-LAT collaboration in the region in Galactic coordinates limited by the intervals $l \in (275, 284)$ and $b \in (-37, -30)$, which contains all SNRs in the main catalogs.

Four of these sources are classified as extended: the larger one, 4FGL J0519.9$-$6845e has a radius of 3\degr\ and essentially includes the most active region of LMC: 57 of the ZMC24 sources are within its area. Two of the other extended sources have a radius of 54\arcmin\ and are in practice contained in the previous one: 4FGL J0530.0$-$6900e,  corresponding to the 30 Doradus complex, and 4FGL J0500.9$-$6945e, which lies in the Far West region. The former region contains six SNRs and the latter five. Furthermore, 4FGL J0531.8-6639e, indicated as LMC-North, has the lowest radius of 36\arcmin, contains one SNR and another one is on its edge.

Four of the remaining point sources within this region are associated with blazars or blazar candidates: 
$i$) 4FGL J0534.1$-$7218 is the likely counterpart of the BL Lac object 5BZB J0533$-$7216: this is the only source not already included in the 4FGL-DR3 version. 
$ii$) 4FGL J0529.3$-$7243, corresponding to a CRATES flat-spectrum source, is located in a rather peripheral position and can actually be a background blazar.
$iii$) 4FGL J0517.9$-$6930c,  with the final code ``c'' that means that it is ``confused with interstellar cloud complexes'',  is within the area of 4FGL J0519.9$-$6845e and is located between the two close SNRs N 120 (9\farcm5) and B0520$-$694 (10\farcm4); it is associated with the source MC4 0518$-$696A; 
$iv$) 4FGL J0535.7$-$6604c is again classified as a bcu and is associated with the radio source PKS 0535$-$66, which is not a radio loud AGN but it is safely identified with the counterpart of the SNR N 63A since early observations (Westerlund \& Mathewson 1966, Dickel et al. 1993); as mentioned above we reported a \gam-ray cluster corresponding to this source in Paper I, well confirmed by a subsequent analysis in Paper II.

Two other sources, namely 4FGL J0537.8$-$6909 (P2 in \citet{ackermann16}) and 4FGL J0524.8$-$6938 (P4), are associated with the two powerful SNRs N 157B and N 132D respectively, and 4FGL J0540.3$-$6920 (P1) is safely identified with the young pulsar PSR J0540$-$6919 \cite{ackermann15a}.
Finally, the two sources 4FGL J0535.2$-$6736 (P3) and 4FGL J0511.4$-$6804 are reported without indication for possible counterparts. \citet{Corbet16} associated the former source with the High-mass X-ray binary CORK053600.0$-$673507, that according to \citet{seward12} appears located in the SNR DEM L241.

In addition to these sources, \citet{eagle2023} reported the detection of a new source in LMC that was associated with the SNR B0453$-$685, classified as pulsar wind nebula (PWN) despite an inner pulsating source has not been detected to now.

\section{Results of cluster analysis at energies above 4 and 10 GeV}    
\label{s:cl410}

\subsection{Results above 10 GeV}  
\label{ss:cl10}

The first step in our analysis was the verification of the SNR detections reported in Paper II. 
It would be expected that the addition of a rather low number of photons (about 25\%) does not change the main findings of that analysis,
however it should be taken into account that even a single photon can modify the cluster parameters and move their values across the selection threshold, or join two close clusters.

As reported in Table 1, all clusters in Paper II are safely confirmed by both methods. 
In particular, the DBSCAN/DENCLUE significance was higher than 3 standard deviations for all sources in the two considered energy ranges, with the exception of the two very close SNRs N~49 and N~49B which were unresolved, while they are separated by MST.
The detection of the SNR N~186D is now confirmed by DBSCAN/DENCLUE at energies higher than 10 GeV, whereas MST analysis gives in this range a cluster with the same number of photons as in Paper II but a slightly lower $M$ due to the higher mean photon density.
In the lower energy range MST finds a cluster of only 9 photons and a marginal $M$ value.

\subsection{Results above 4 GeV}   
\label{ss:cl4}

As explained above, the search for photon clusters above 4 GeV is more difficult because either of the large increase of the background and the broad PSF with the consequent decrease in the density contrast.
As a consequence, the definition of selection thresholds and the significance of clusters cannot be safely established. 
Therefore, we first applied the DBSCAN /DENCLUE algorithm whose output gives significance estimates of individual clusters, and then verified them by means of MST.
This analysis produced a sample of 35 clusters with a significance higher than 3 standard deviations, 12 of which already found in Paper II. 
Seven of the remaining clusters were found with a close positional correspondence with SNRs, well within the PSF radius at this energy. 
The properties and the search for counterparts of the remaining 16 clusters will be presented in another paper.

Note that the values of the magnitude $M$, our main selection parameter 
defined in Eq.(2), are generally higher than 20 that is the threshold 
adopted in the cluster search above 10 GeV that can also be adopted at 
energies higher than 4 GeV, where magnitude values of clusters associated
with sources in the 4FGL-DR4 catalog are even lower.
As reported in Table 2, only two clusters have $M$ lower than 20: one is 
19.8, very close to the threshold value, and the other has $M$ = 18.3 that 
is well acceptable at these energies.

We also performed a new MST analysis, with $\Lambda_{cut}$ = 0:8, in a 
different region defined by the following boundaries $l \in$ (274; 281) 
and $b \in$ ($-$40; $-$33), thus excluding the bright 30 Dor complex. 
The results of this search confirmed the clusters corresponding to the 
powerful SNRs with the same numbers of photons but with slightly
higher magnitude values above the nominal threshold of 20, namely:
$M =$ 20.2 for B0453$-$685, and $M =$ 21.8 for N103B, and 29.4 for B0519$-$690. 
Just for comparison, we found two clusters corresponding to non-extended 
sources in 4FGL-DR4: the cluster associated with 4FGL J0511.4$-$6804 has 
18 photons and $M =$ 43.4, and the one associated with 4FGL J0443.3$-$6652 has
10 photons and $M =$ 21.2, comparable to those of the three remnants indicated 
above. 
As a further verification about SNR N103B, we analyzed the data at energies 
higher than 6 GeV and found very close to its position a cluster of only 5 
photons but a quite high clustering parameter $g =$ 3.2 as defined in 
Sect. 3.2.

Five clusters were found in the standard analysis over the very large field; the cluster likely associated with \lbrack HP99\rbrack 1139 was found at energies higher than 6 GeV but not above 4 GeV, and the one associated with the complex DEM L316A/B was not found in the very large field whereas a positive detection resulted in a narrow field excluding the bright region of 30 Dor. 
We also searched in the primary selection list of clusters above 10 GeV for entries corresponding to these remnants and found only a low significance cluster at the position of DEM L241, whose photon number is the highest.
The properties of these two interesting sources will be discussed in detail in the following Sections~\ref{ss:cld316} and ~\ref{ss:cld241}.

In Figure 1 a map of the galaxy is shown with the locations of clusters: the new ones are outside the very bright region of 30 Dor, in particular the four clusters associated with faint SNRs are placed at distances greater that 1 degree from the external boundary.
The other three clusters in Table 2, i.e., B0453$-$685, N~103B, B0519$-$690, are in the subsample of bright X-ray emitters undetected in Paper II, and therefore they could be considered as best candidates; in particular, the first of them was already detected by \citet{eagle2023}, while the other two are new findings.

\subsection{Properties of the cluster associated with DEM L316A/B}    
\label{ss:cld316}

One of the new clusters detected at energies higher than 4 GeV is spatially close to the peculiar object DEM L316A/B which can be considered a likely counterpart.
This source appears composed by two overlapping remnants (indicated by the letters A and B), investigated in the radio and optical by \citet{williams97}, however the possibility that the two remnants are actually interacting is not confirmed by X-ray data analyses \citep{williamschu05,toledo09}.
The two remnants are different in size but both have rather large diameters (2\arcmin and 3\arcmin, corresponding to about 30 and 40 pc) and are relatively evolved with estimated ages of 27 kyr and 39 kyr.
The centroid coordinates of the cluster given in Table 2 are closer to the remnant A, but it is reasonable that both these objects can contribute to the photon cluster, because $R_\mathrm{max}$ is equal to 6\farcm7 and the angular distance between the centers of the remnants is around 2\farcm4, lower than the instrumental PSF at 4 GeV. 
The spectra of the two remnants are remarkably different, in particular for the Fe abundance, which indicates that the A shell had a Ia supernova origin, whereas the B originated from a core-collapse (CC) supernova \citep{nishiuchi01}.
Their X-ray luminosities in the \citet{maggi16} catalog are remarkably similar.

The \gam-ray cluster is not very significant in the MST search in large fields and an acceptable value of $M$ is obtained only using a narrow field excluding the 30 Dor region.
Similar analyses at energies higher than 6 or 10 GeV gave clusters with a lower $M$ values below the threshold.
Thus spectrum of this cluster should be very steep or exhibit a rather sharp cut-off just above 4 GeV. 

\subsection{Properties of the cluster associated with DEM L241}    
\label{ss:cld241}

The source P3 in the early paper by \citet{ackermann16} was reported as ``unassociated'' since DEM L241, as well other possible counterparts, were at a distance from the best fit position of about 11\arcmin\ at the 2$\sigma$ limit. 
However, its spectrum was found to be quite soft with an estimated photon index equal to 2.8.
The corresponding source in the most recent \fl catalog, 4FGL J0535.2-6736, is at angular distance of about 5\arcmin\ from the remnant and the counterpart is the HMXB mentioned in Section~\ref{s:clF4}.
This association appears safe given that the \gam-ray flux exhibits a very significant modulation with a period of 10.3 days, the same as the HMXB \citep{Corbet16}.
An X-ray source was discovered by \citet{seward12} inside the remnant in a \emph{ Chandra} image of DEM L241; spectral data suggest that it is comprising an O-type star and a neutron star.

In Paper I a cluster of only 4 photons above 10 GeV with a low $g$ was found at an angular distance of 2\farcm8 from the remnant. 
A further analysis at energies higher than 3 GeV showed evidence for a richer cluster, in agreement with the softness of its spectrum.
In Paper II we found at energies    higher than 6 GeV the cluster MST(83.643, $-$67.326) that, however, was not close enough to the 4FGL-DR4 source for a safe association.
The present analysis at energies higher than 4 GeV provided a very significant cluster either with DBSCAN/DENCLUE and with MST (25 photons, $M$ = 56.67) at a position fully consistent with the known source and the SNR.
The search in the primary MST selection sample above 10 GeV gives a poorly significant cluster at (83.8996, $-$67.5959) with 5 photons and $M = 9.6$, consistent again with a soft spectral distribution.
One can therefore conclude that high energy photons are mostly emitted by the HMXB rather than the remnant, and consequently this source is not included in the sample of genuine \gam-ray loud SNR.

\section{Discussion and Conclusions}

\label{s:conc}

The present analysis, exploiting 15 years of \fl observations, confirms that the LMC region exhibits a significant clumpy structure in the \gam-ray emission, with several photon clusters closely associated with SNRs. 
Our results confirm all previous detections and found positive indications for at least 8 new clusters with a spatial correspondence with other SNRs, increasing thus the number of remnants in LMC candidate or detected in the high energy \gam-rays to 16 sources, one of which was already found associated with a
X-ray binary in the remnant DEM L241. 
The number of \gam-ray-detected SNRs in the LMC above 4 GeV is now increased to 15 and represents approximately 25\% of the \citet{maggi16} sample or 19\% of the eROSITA catalog. 

The detected clusters provide valuable insights into the interplay between the intrinsic properties of SNRs and their \gam-ray emission. A clear trend emerges, linking higher X-ray luminosity in the $0.3-8$ keV band ($L_{\rm X} > 10^{36}$~erg~s$^{-1}$; a general indicator of the power output of a SNR) with a higher probability of \gam-ray detection. Among the 12 brightest SNRs detected in X-rays, nine are present in our \gam-ray cluster list. This correlation underscores the role of the environment, where shock interaction with dense interstellar material and particle acceleration processes produce detectable \gam-ray signatures. The remaining six detected SNRs have X-ray luminosities in the range $0.5-4.7 \times 10^{35}$~erg~s$^{-1}$ as many other sources in the sample.

However, some challenges remain: ($i$) three SNRs in the bright subsample (B0509$-$67.5, DEM~L71, and N~23) are not associated with significant \gam-ray clusters, despite expectations of detectable emission; ($ii$) six low-luminosity SNRs are detected in \gam-rays, with different degrees of significance, while several other sources of comparable luminosity remain undetected.

Concerning the first point, the properties of the three remnants suggest plausible explanations for their lack of \gam-ray detection. B0509$-$67.5 and DEM~L71 both exhibit low cosmic-ray acceleration efficiencies. The remnant DEM~L71 exhibits enhanced Fe emission and is classified as a Type Ia SNR; its age is estimated to be approximately 6700 years \citep{frank19, bilir22}. According to \citet{rakowski03} and \citet{ghavamian03}, the absence of a non-thermal X-ray component and the low radio flux suggest that the blast wave energy loss to cosmic-ray acceleration is minimal at the current epoch, reducing the probability to detect high-energy \gam-ray emission. The remnant B0509$-$67.5, on the other hand, is a very young Type Ia SNR, originating from an explosion about 400 years ago \citep{yamag14, kosenko14}. Based on optical spectroscopic studies, \citet{hovey18} determined that this SNR also has a low efficiency for cosmic-ray acceleration. This inefficiency likely explains its lack of significant \gam-ray emission, despite its young age and high-energy potential. From these observational findings, we conclude that the low cosmic-ray acceleration efficiencies of these remnants can be responsible for their weak or absent high-energy \gam-ray signatures.

The case of N~23 is more complex. This is classified as a CC SNR and contains a candidate PWN identified as the point-like X-ray source CXOU J050552.3$-$680141 \citep{hayato06}. Despite its classification and the potential PWN contribution, N~23 exhibits only faint nebular X-ray emission, and its \gam-ray emission remains undetected. This suggests that other factors, such as the properties of the surrounding medium or a lack of sufficient particle acceleration mechanisms, could be suppressing its high-energy output. Further multi-wavelength studies, particularly in the radio and X-ray bands, are needed to clarify the mechanisms limiting \gam-ray production in N~23.

Interestingly, these three SNRs are located in a relatively compact region of the LMC, distinct from the high-activity 30 Dor region. The angular separation between DEM~L71 and N~23 is only 9$'$.2, corresponding to a projected distance of approximately 134 pc, while B0509$-$67.5 lies about 37$'$ away, or $\sim 540$ pc. This spatial clustering raises the possibility of a shared local environment or regional conditions affecting their \gam-ray emission, which merits further exploration.

For what concerns the low-luminosity SNRs associated with \gam-ray clusters, the disparity in \gam-ray detection between these six remnants and the remaining low-luminosity SNRs of the sample may stem from environmental or intrinsic factors. These include variations in shock efficiency, differences in the density and structure of the surrounding medium, or the strength and configuration of magnetic fields. Such factors significantly affect the ability of remnants to accelerate particles to relativistic energies and produce detectable \gam-ray emission. Further investigation is essential to clarify why only six out of about 40 remnants in this category exhibit detectable \gam-ray emission, potentially shedding light on the mechanisms driving their high-energy output.

The properties of three of these six SNRs were already discussed in Paper II. 
N~44 is a complex nebular system that includes three components, one of which is the SNR 0523$-$679. 
The source B0528$-$692, now confirmed as an SNR through the new analysis, shows increased significance and a reduced angular distance between the cluster centroid and the SNR coordinates. However, its position within the brightest region of the LMC raises the possibility of contamination from nearby sources. 
For the third remnant, N~186D, the photon count remains consistent with Paper II, but its lower $M$-value results in no significant improvement in detection confidence. 
Despite this, the association between the cluster and this remnant remains robust.

Three low luminosity SNRs are detected above 4 GeV: the case of DEM L316A/B is discussed in Sect.
~\ref{ss:cld316} while literature data on the other two remnants are rather poor.
An X-ray image, showing a central bright emission, and a spectral analysis of \lbrack HP99\rbrack 1139 was reported by \citet{whelan14}, who classified it as Ia type with an age around 19 kyr and a large size which makes it one of the most extended remnants in the LMC.
The detection of a middle-aged SNR below 10 GeV is somehow in line with what happens in our Galaxy, where evolved SNRs are typically characterized by a soft (likely hadronic) $\gamma-$ray emission, barely extending beyond 10 GeV, while the spectral energy distribution of younger SNRs peaks at higher energies \citep[e.g.][]{Funk2015}.
B0532$-$675 has been investigated by \citet{li22} who did not observe an optical shell pattern but its extension can be derived from the non-thermal radio emission.
The OB association LH75 (NGC 2011) is projected within the remnant boundaries and they propose that the supernova progenitor was a star of about 15 solar masses in this association and exploded in a HI cavity.
The presence of an OB association might be a possibility to enhance high energy emission as proposed by \citet{kava19} and \citet{aron24} for TeV sources detected in LMC, but in our case this require more focused studies.

The results of new cluster analyses of the photon field of the LMC region at
TeV energies, observed by the Fermi-LAT experiment in 15 years, can be 
summarized as follows:
\begin{itemize}
    \item We detected photon clusters whose centroids are at angular separations of only few arcminutes from 15 SNRs: more specifically, 8 SNR are found above 10 GeV, 1
    above 6 GeV and the other 6 remnants above 4 GeV.
    \item One of these clusters corresponds to a very close pair of SNRs (DEM L316A and DEM L316B) having a separation smaller than the PSF diameter and therefore
    their \gam-ray emission cannot be resolved.
    \item There is robust evidence for a cluster correspondent to the SNR DEM L241,   a source already detected and exhibiting a clear modulation with the period of an X-ray binary inside the remnant.
    \item We also searched for clusters close to the position of the very young SNR 1987A and did not find any significant feature. We recall, however, that this source is close to the bright region of 30 Dor and the local high spatial density of photons makes quite hard the cluster extraction.
    \item 
    The remnants previously reported in Paper II were all of the Core-Collapse type,
    while in the present analysis, we also found two Ia SNRs although only at energies 
    higher than 4 GeV. Their emission, therefore, appears generally softer than CC remnants. It is important to verify this trend with other sources but these investigations require a much longer exposure to achieve sufficient photon statistics.
    
\end{itemize}
Given the proximity, and, particularly the low metallicity and an intense star formation history, the LMC provides a unique laboratory with conditions similar to those of primeval galaxies in the early universe. Our findings are therefore useful to refine the understanding of young SNR evolution, particle acceleration, and progenitor-type differences, emphasizing the need for deeper high-energy observations to verify spectral trends and detect faint sources.

\begin{acknowledgements}
F.B., M.M., and S.O. acknowledge financial contribution from the PRIN 2022 (20224MNC5A) - ``Life, death and after-death of massive stars'' funded by European Union – Next Generation EU, and the INAF Theory Grant ``Supernova remnants as probes for the structure and mass-loss history of the progenitor systems''.
\end{acknowledgements}

\bibliographystyle{aa}
\bibliography{bibliography} 

\begin{thebibliography}{55}
\expandafter\ifx\csname natexlab\endcsname\relax\def\natexlab#1{#1}\fi

\bibitem[{{Abdo} {et~al.}(2010){Abdo}, {Ackermann}, {Ajello}, {Atwood}, {Baldini}, {Ballet}, {Barbiellini}, {Bastieri}, {Baughman}, {Bechtol}, {Bellazzini}, {Berenji}, {Blandford}, {Bloom}, {Bonamente}, {Borgland}, {Bregeon}, {Brez}, {Brigida}, {Bruel}, {Burnett}, {Buson}, {Caliandro}, {Cameron}, {Caraveo}, {Casandjian}, {Cecchi}, {{\c{C}}elik}, {Chekhtman}, {Cheung}, {Chiang}, {Ciprini}, {Claus}, {Cohen-Tanugi}, {Cominsky}, {Conrad}, {Cutini}, {Dermer}, {de Angelis}, {de Palma}, {Digel}, {Silva}, {Drell}, {Dubois}, {Dumora}, {Farnier}, {Favuzzi}, {Fegan}, {Focke}, {Fortin}, {Frailis}, {Fukazawa}, {Fusco}, {Gargano}, {Gasparrini}, {Gehrels}, {Germani}, {Giavitto}, {Giebels}, {Giglietto}, {Giordano}, {Glanzman}, {Godfrey}, {Gotthelf}, {Grenier}, {Grondin}, {Grove}, {Guillemot}, {Guiriec}, {Hanabata}, {Harding}, {Hayashida}, {Hays}, {Horan}, {Hughes}, {Jackson}, {Jean}, {J{\'o}hannesson}, {Johnson}, {Johnson}, {Johnson}, {Johnson}, {Kamae}, {Katagiri}, {Kataoka}, {Kawai}, {Kerr}, {Kn{\"o}dlseder}, {Kocian}, {Kuss}, {Lande}, {Latronico}, {Lemoine-Goumard}, {Longo}, {Loparco}, {Lott}, {Lovellette}, {Lubrano}, {Madejski}, {Makeev}, {Marshall}, {Martin}, {Mazziotta}, {McConville}, {McEnery}, {Meurer}, {Michelson}, {Mitthumsiri}, {Mizuno}, {Moiseev}, {Monte}, {Monzani}, {Morselli}, {Moskalenko}, {Murgia}, {Nolan}, {Norris}, {Nuss}, {Ohsugi}, {Omodei}, {Orlando}, {Ormes}, {Paneque}, {Parent}, {Pelassa}, {Pepe}, {Pesce-Rollins}, {Piron}, {Porter}, {Rain{\`o}}, {Rando}, {Razzano}, {Reimer}, {Reimer}, {Reposeur}, {Ritz}, {Rodriguez}, {Romani}, {Roth}, {Ryde}, {Sadrozinski}, {Sanchez}, {Sander}, {Saz Parkinson}, {Scargle}, {Sellerholm}, {Sgr{\`o}}, {Siskind}, {Smith}, {Smith}, {Spandre}, {Spinelli}, {Starck}, {Strickman}, {Strong}, {Suson}, {Tajima}, {Takahashi}, {Tanaka}, {Thayer}, {Thayer}, {Thompson}, {Tibaldo}, {Torres}, {Tosti}, {Tramacere}, {Uchiyama}, {Usher}, {Vasileiou}, {Venter}, {Vilchez}, {Vitale}, {Waite}, {Wang}, {Weltevrede}, {Winer}, {Wood}, {Ylinen}, \& {Ziegler}}]{abdo10}
{Abdo}, A.~A., {Ackermann}, M., {Ajello}, M., {et~al.} 2010, \aap, 512, A7

\bibitem[{{Abdollahi} {et~al.}(2020){Abdollahi}, {Acero}, {Ackermann}, {Ajello}, {Atwood}, {Axelsson}, {Baldini}, {Ballet}, {Barbiellini}, {Bastieri}, {Becerra Gonzalez}, {Bellazzini}, {Berretta}, {Bissaldi}, {Blandford}, {Bloom}, {Bonino}, {Bottacini}, {Brandt}, {Bregeon}, {Bruel}, {Buehler}, {Burnett}, {Buson}, {Cameron}, {Caputo}, {Caraveo}, {Casandjian}, {Castro}, {Cavazzuti}, {Charles}, {Chaty}, {Chen}, {Cheung}, {Chiaro}, {Ciprini}, {Cohen-Tanugi}, {Cominsky}, {Coronado-Bl{\'a}zquez}, {Costantin}, {Cuoco}, {Cutini}, {D'Ammando}, {DeKlotz}, {de la Torre Luque}, {de Palma}, {Desai}, {Digel}, {Di Lalla}, {Di Mauro}, {Di Venere}, {Dom{\'\i}nguez}, {Dumora}, {Fana Dirirsa}, {Fegan}, {Ferrara}, {Franckowiak}, {Fukazawa}, {Funk}, {Fusco}, {Gargano}, {Gasparrini}, {Giglietto}, {Giommi}, {Giordano}, {Giroletti}, {Glanzman}, {Green}, {Grenier}, {Griffin}, {Grondin}, {Grove}, {Guiriec}, {Harding}, {Hayashi}, {Hays}, {Hewitt}, {Horan}, {J{\'o}hannesson}, {Johnson}, {Kamae}, {Kerr}, {Kocevski}, {Kovac'evic'}, {Kuss}, {Landriu}, {Larsson}, {Latronico}, {Lemoine-Goumard}, {Li}, {Liodakis}, {Longo}, {Loparco}, {Lott}, {Lovellette}, {Lubrano}, {Madejski}, {Maldera}, {Malyshev}, {Manfreda}, {Marchesini}, {Marcotulli}, {Mart{\'\i}-Devesa}, {Martin}, {Massaro}, {Mazziotta}, {McEnery}, {Mereu}, {Meyer}, {Michelson}, {Mirabal}, {Mizuno}, {Monzani}, {Morselli}, {Moskalenko}, {Negro}, {Nuss}, {Ojha}, {Omodei}, {Orienti}, {Orlando}, {Ormes}, {Palatiello}, {Paliya}, {Paneque}, {Pei}, {Pe{\~n}a-Herazo}, {Perkins}, {Persic}, {Pesce-Rollins}, {Petrosian}, {Petrov}, {Piron}, {Poon}, {Porter}, {Principe}, {Rain{\`o}}, {Rando}, {Razzano}, {Razzaque}, {Reimer}, {Reimer}, {Remy}, {Reposeur}, {Romani}, {Saz Parkinson}, {Schinzel}, {Serini}, {Sgr{\`o}}, {Siskind}, {Smith}, {Spandre}, {Spinelli}, {Strong}, {Suson}, {Tajima}, {Takahashi}, {Tak}, {Thayer}, {Thompson}, {Tibaldo}, {Torres}, {Torresi}, {Valverde}, {Van Klaveren}, {van Zyl}, {Wood}, {Yassine}, \& {Zaharijas}}]{abdollahi20}
{Abdollahi}, S., {Acero}, F., {Ackermann}, M., {et~al.} 2020, \apjs, 247, 33

\bibitem[{{Ackermann} {et~al.}(2012){Ackermann}, {Ajello}, {Albert}, {Allafort}, {Atwood}, {Axelsson}, {Baldini}, {Ballet}, {et~al.}}]{ackermann12}
{Ackermann}, M., {Ajello}, M., {Albert}, A., {et~al.} 2012, \apjs, 203, 4

\bibitem[{{Ackermann} {et~al.}(2016){Ackermann}, {Albert}, {Atwood}, {Baldini}, {Ballet}, {Barbiellini}, {Bastieri}, {Bellazzini}, {Bissaldi}, {et~al.}}]{ackermann16}
{Ackermann}, M., {Albert}, A., {Atwood}, W.~B., {et~al.} 2016, \aap, 586, A71

\bibitem[{{Aharonian} {et~al.}(2024){Aharonian}, {Benkhali}, {Aschersleben}, {Ashkar}, {Backes}, {Martins}, {Batzofin}, {Becherini}, {Berge}, {Bernl{\"o}hr}, {B{\"o}ttcher}, {Bolmont}, {de Bony de Lavergne}, {Borowska}, {Brose}, {Brown}, {Brun}, {Bruno}, {Burger-Scheidlin}, {Casanova}, {Celic}, {Cerruti}, {Chand}, {Chandra}, {Chen}, {Chibueze}, {Chibueze}, {Cotter}, {Cristofari}, {Devin}, {Djannati-Ata{\"\i}}, {Djuvsland}, {Dmytriiev}, {Egberts}, {Einecke}, {Feijen}, {Filipovic}, {Fontaine}, {Funk}, {Gabici}, {Gallant}, {Glicenstein}, {Glombitza}, {Grolleron}, {Haerer}, {He{\ss}}, {Hinton}, {Hofmann}, {Holch}, {Horns}, {Huang}, {Jamrozy}, {Jankowsky}, {Jung-Richardt}, {Kasai}, {Katarzy{\'n}ski}, {Khatoon}, {Kh{\'e}lifi}, {Klu{\'z}niak}, {Komin}, {Kosack}, {Kostunin}, {Kundu}, {Lang}, {Le Stum}, {Lemi{\`e}re}, {Lemoine-Goumard}, {Lenain}, {Leuschner}, {Mackey}, {Marandon}, {Mart{\'\i}-Devesa}, {Marx}, {Mehta}, {Mitchell}, {Moderski}, {Moghadam}, {Mohrmann}, {Montanari}, {Moulin}, {de Naurois}, {Niemiec}, {Ohm}, {Olivera-Nieto}, {de Ona Wilhelmi}, {Ostrowski}, {Panny}, {Pensec}, {Peron}, {P{\"u}hlhofer}, {Quirrenbach}, {Ravikularaman}, {Regeard}, {Reimer}, {Reimer}, {Ren}, {Renaud}, {Reville}, {Rieger}, {Rowell}, {Rudak}, {Ruiz-Velasco}, {Sabri}, {Sahakian}, {Salzmann}, {Santangelo}, {Sasaki}, {Sch{\"a}fer}, {Sch{\"u}ssler}, {Schutte}, {Sol}, {Spencer}, {Stawarz}, {Steinmassl}, {Steppa}, {Streil}, {Sushch}, {Taylor}, {Terrier}, {Tsirou}, {Tsuji}, {van Eldik}, {Vecchi}, {Venter}, {Vink}, {Wagner}, {White}, {Wierzcholska}, {Zacharias}, {Zdziarski}, {Zech}, {{\.Z}ywucka}, \& {H.~E.~S.~S. Collaboration}}]{aron24}
{Aharonian}, F., {Benkhali}, F.~A., {Aschersleben}, J., {et~al.} 2024, \apjl, 970, L21

\bibitem[{{Ajello} {et~al.}(2017){Ajello}, {Atwood}, {Baldini}, {Ballet}, {Barbiellini}, {Bastieri}, {Bellazzini}, {Bissaldi}, {Blandford}, {Bloom}, {Bonino}, {Bregeon}, {Britto}, {Bruel}, {Buehler}, {Buson}, {Cameron}, {Caputo}, {Caragiulo}, {Caraveo}, {Cavazzuti}, {Cecchi}, {Charles}, {Chekhtman}, {Cheung}, {Chiaro}, {Ciprini}, {Cohen}, {Costantin}, {Costanza}, {Cuoco}, {Cutini}, {D'Ammando}, {de Palma}, {Desiante}, {Digel}, {Di Lalla}, {Di Mauro}, {Di Venere}, {Dom{\'\i}nguez}, {Drell}, {Dumora}, {Favuzzi}, {Fegan}, {Ferrara}, {Fortin}, {Franckowiak}, {Fukazawa}, {Funk}, {Fusco}, {Gargano}, {Gasparrini}, {Giglietto}, {Giommi}, {Giordano}, {Giroletti}, {Glanzman}, {Green}, {Grenier}, {Grondin}, {Grove}, {Guillemot}, {Guiriec}, {Harding}, {Hays}, {Hewitt}, {Horan}, {J{\'o}hannesson}, {Kensei}, {Kuss}, {La Mura}, {Larsson}, {Latronico}, {Lemoine-Goumard}, {Li}, {Longo}, {Loparco}, {Lott}, {Lubrano}, {Magill}, {Maldera}, {Manfreda}, {Mazziotta}, {McEnery}, {Meyer}, {Michelson}, {Mirabal}, {Mitthumsiri}, {Mizuno}, {Moiseev}, {Monzani}, {Morselli}, {Moskalenko}, {Negro}, {Nuss}, {Ohsugi}, {Omodei}, {Orienti}, {Orlando}, {Palatiello}, {Paliya}, {Paneque}, {Perkins}, {Persic}, {Pesce-Rollins}, {Piron}, {Porter}, {Principe}, {Rain{\`o}}, {Rando}, {Razzano}, {Razzaque}, {Reimer}, {Reimer}, {Reposeur}, {Saz Parkinson}, {Sgr{\`o}}, {Simone}, {Siskind}, {Spada}, {Spandre}, {Spinelli}, {Stawarz}, {Suson}, {Takahashi}, {Tak}, {Thayer}, {Thayer}, {Thompson}, {Torres}, {Torresi}, {Troja}, {Vianello}, {Wood}, \& {Wood}}]{ajello17}
{Ajello}, M., {Atwood}, W.~B., {Baldini}, L., {et~al.} 2017, \apjs, 232, 18

\bibitem[{{Alan} \& {Bilir}(2022)}]{bilir22}
{Alan}, N. \& {Bilir}, S. 2022, \mnras, 511, 5018

\bibitem[{{Ballet} {et~al.}(2020){Ballet}, {Burnett}, {Digel}, \& {Lott}}]{ballet20}
{Ballet}, J., {Burnett}, T.~H., {Digel}, S.~W., \& {Lott}, B. 2020, arXiv e-prints, arXiv:2005.11208

\bibitem[{{Bernieri} {et~al.}(2013){Bernieri}, {Campana}, {Massaro}, {Paggi}, \& {Tramacere}}]{bernieri13}
{Bernieri}, E., {Campana}, R., {Massaro}, E., {Paggi}, A., \& {Tramacere}, A. 2013, \aap, 551, L5

\bibitem[{{Bozzetto} {et~al.}(2017){Bozzetto}, {Filipovi{\'c}}, {Vukoti{\'c}}, {Pavlovi{\'c}}, {Uro{\v s}evi{\'c}}, {Kavanagh}, {Arbutina}, {Maggi}, {Sasaki}, {Haberl}, {Crawford}, {Roper}, {Grieve}, \& {Points}}]{bozzetto17}
{Bozzetto}, L.~M., {Filipovi{\'c}}, M.~D., {Vukoti{\'c}}, B., {et~al.} 2017, \apjs, 230, 2

\bibitem[{{Campana} {et~al.}(2013){Campana}, {Bernieri}, {Massaro}, {Tinebra}, \& {Tosti}}]{campana13}
{Campana}, R., {Bernieri}, E., {Massaro}, E., {Tinebra}, F., \& {Tosti}, G. 2013, \apss, 347, 169

\bibitem[{{Campana} {et~al.}(2017){Campana}, {Maselli}, {Bernieri}, \& {Massaro}}]{campana17}
{Campana}, R., {Maselli}, A., {Bernieri}, E., \& {Massaro}, E. 2017, \mnras, 465, 2784

\bibitem[{{Campana} \& {Massaro}(2021)}]{campana21}
{Campana}, R. \& {Massaro}, E. 2021, \aap, 652, A6

\bibitem[{{Campana} {et~al.}(2016{\natexlab{a}}){Campana}, {Massaro}, \& {Bernieri}}]{paperIV}
{Campana}, R., {Massaro}, E., \& {Bernieri}, E. 2016{\natexlab{a}}, \apss, 361, 367

\bibitem[{{Campana} {et~al.}(2016{\natexlab{b}}){Campana}, {Massaro}, \& {Bernieri}}]{paperIII}
{Campana}, R., {Massaro}, E., \& {Bernieri}, E. 2016{\natexlab{b}}, \apss, 361, 185

\bibitem[{{Campana} {et~al.}(2016{\natexlab{c}}){Campana}, {Massaro}, \& {Bernieri}}]{paperII}
{Campana}, R., {Massaro}, E., \& {Bernieri}, E. 2016{\natexlab{c}}, \apss, 361, 183

\bibitem[{{Campana} {et~al.}(2018{\natexlab{a}}){Campana}, {Massaro}, \& {Bernieri}}]{campana18b}
{Campana}, R., {Massaro}, E., \& {Bernieri}, E. 2018{\natexlab{a}}, \aap, 619, A23

\bibitem[{{Campana} {et~al.}(2018{\natexlab{b}}){Campana}, {Massaro}, \& {Bernieri}}]{campana18}
{Campana}, R., {Massaro}, E., \& {Bernieri}, E. 2018{\natexlab{b}}, \apss, 363, 144

\bibitem[{{Campana} {et~al.}(2015){Campana}, {Massaro}, {Bernieri}, \& {D'Amato}}]{paperI}
{Campana}, R., {Massaro}, E., {Bernieri}, E., \& {D'Amato}, Q. 2015, \apss, 360, 19

\bibitem[{{Campana} {et~al.}(2022){Campana}, {Massaro}, {Bocchino}, {Miceli}, {Orlando}, \& {Tramacere}}]{campana22}
{Campana}, R., {Massaro}, E., {Bocchino}, F., {et~al.} 2022, \mnras, 515, 1676

\bibitem[{{Campana} {et~al.}(2008){Campana}, {Massaro}, {Gasparrini}, {Cutini}, \& {Tramacere}}]{campana08}
{Campana}, R., {Massaro}, E., {Gasparrini}, D., {Cutini}, S., \& {Tramacere}, A. 2008, \mnras, 383, 1166

\bibitem[{{Corbet} {et~al.}(2016){Corbet}, {Chomiuk}, {Coe}, {Coley}, {Dubus}, {Edwards}, {Martin}, {McBride}, {Stevens}, {Strader}, {Townsend}, \& {Udalski}}]{Corbet16}
{Corbet}, R.~H.~D., {Chomiuk}, L., {Coe}, M.~J., {et~al.} 2016, \apj, 829, 105

\bibitem[{{Cormen} {et~al.}(2009){Cormen}, {Leiserson}, Rivest, \& {Stein}}]{cormen09}
{Cormen}, T., {Leiserson}, C., Rivest, R., \& {Stein}, C. 2009, Introduction to Algorithms, 3rd edn. (Cambridge, USA: MIT Press)

\bibitem[{{Cusumano} {et~al.}(1998){Cusumano}, {Maccarone}, {Mineo}, {Sacco}, {Massaro}, {Bandiera}, \& {Salvati}}]{cusumano98}
{Cusumano}, G., {Maccarone}, M.~C., {Mineo}, T., {et~al.} 1998, \aap, 333, L55

\bibitem[{{Eagle} {et~al.}(2023){Eagle}, {Castro}, {Mahhov}, {Gelfand}, {Kerr}, {Slane}, {Ballet}, {Acero}, {Straal}, \& {Ajello}}]{eagle2023}
{Eagle}, J., {Castro}, D., {Mahhov}, P., {et~al.} 2023, \apj, 945, 4

\bibitem[{{Fermi LAT Collaboration} {et~al.}(2015){Fermi LAT Collaboration}, {Ackermann}, {Albert}, {Baldini}, {Ballet}, {Barbiellini}, {Barbieri}, {Bastieri}, {Bellazzini}, {et~al.}}]{ackermann15a}
{Fermi LAT Collaboration}, {Ackermann}, M., {Albert}, A., {et~al.} 2015, Science, 350, 801

\bibitem[{{Frank} {et~al.}(2019){Frank}, {Dwarkadas}, {Panfichi}, {Crum}, \& {Burrows}}]{frank19}
{Frank}, K.~A., {Dwarkadas}, V., {Panfichi}, A., {Crum}, R.~M., \& {Burrows}, D.~N. 2019, \apj, 875, 14

\bibitem[{{Funk}(2015)}]{Funk2015}
{Funk}, S. 2015, Annual Review of Nuclear and Particle Science, 65, 245

\bibitem[{{Ghavamian} {et~al.}(2003){Ghavamian}, {Rakowski}, {Hughes}, \& {Williams}}]{ghavamian03}
{Ghavamian}, P., {Rakowski}, C.~E., {Hughes}, J.~P., \& {Williams}, T.~B. 2003, \apj, 590, 833

\bibitem[{{Hayato} {et~al.}(2006){Hayato}, {Bamba}, {Tamagawa}, \& {Kawabata}}]{hayato06}
{Hayato}, A., {Bamba}, A., {Tamagawa}, T., \& {Kawabata}, K. 2006, \apj, 653, 280

\bibitem[{{H.E.S.S.~Collaboration} {et~al.}(2012){H.E.S.S.~Collaboration}, {Abramowski}, {Acero}, {Aharonian}, {Akhperjanian}, {Anton}, {Balenderan}, {Balzer}, {Barnacka}, {et~al.}}]{abramowski12}
{H.E.S.S.~Collaboration}, {Abramowski}, A., {Acero}, F., {et~al.} 2012, \aap, 545, L2

\bibitem[{{H.E.S.S.~Collaboration} {et~al.}(2015){H.E.S.S.~Collaboration}, {Abramowski}, {Aharonian}, {Ait Benkhali}, {Akhperjanian}, {Ang{\"u}ner}, {Backes}, {Balenderan}, {Balzer}, {Barnacka}, \& et~al.}]{abramowski15}
{H.E.S.S.~Collaboration}, {Abramowski}, A., {Aharonian}, F., {et~al.} 2015, Science, 347, 406

\bibitem[{Hinneburg \& Gabriel(2007)}]{Hinneburg:2007tq}
Hinneburg, A. \& Gabriel, H.-H. 2007, 70

\bibitem[{Hinneburg \& Keim(1998)}]{Hinneburg:1998wo}
Hinneburg, A. \& Keim, D.~A. 1998, 98, 58

\bibitem[{{Hovey} {et~al.}(2018){Hovey}, {Hughes}, {McCully}, {Pandya}, \& {Eriksen}}]{hovey18}
{Hovey}, L., {Hughes}, J.~P., {McCully}, C., {Pandya}, V., \& {Eriksen}, K. 2018, \apj, 862, 148

\bibitem[{{Kavanagh} {et~al.}(2019){Kavanagh}, {Vink}, {Sasaki}, {Chu}, {Filipovi{\'c}}, {Ohm}, {Haberl}, {Manojlovic}, \& {Maggi}}]{kava19}
{Kavanagh}, P.~J., {Vink}, J., {Sasaki}, M., {et~al.} 2019, \aap, 621, A138

\bibitem[{{Kosenko} {et~al.}(2014){Kosenko}, {Ferrand}, \& {Decourchelle}}]{kosenko14}
{Kosenko}, D., {Ferrand}, G., \& {Decourchelle}, A. 2014, \mnras, 443, 1390

\bibitem[{{Li} {et~al.}(2022){Li}, {Chu}, {Chuang}, \& {Li}}]{li22}
{Li}, C.-J., {Chu}, Y.-H., {Chuang}, C.-Y., \& {Li}, G.-H. 2022, \aj, 163, 30

\bibitem[{{Li} \& {Ma}(1983)}]{Li&Ma}
{Li}, T.~P. \& {Ma}, Y.~Q. 1983, \apj, 272, 317

\bibitem[{{Maggi} {et~al.}(2016){Maggi}, {Haberl}, {Kavanagh}, {Sasaki}, {Bozzetto}, {Filipovi{\'c}}, {Vasilopoulos}, {Pietsch}, {Points}, {Chu}, {Dickel}, {Ehle}, {Williams}, \& {Greiner}}]{maggi16}
{Maggi}, P., {Haberl}, F., {Kavanagh}, P.~J., {et~al.} 2016, \aap, 585, A162

\bibitem[{{Marshall} {et~al.}(1998){Marshall}, {Gotthelf}, {Zhang}, {Middleditch}, \& {Wang}}]{marshall98}
{Marshall}, F.~E., {Gotthelf}, E.~V., {Zhang}, W., {Middleditch}, J., \& {Wang}, Q.~D. 1998, \apjl, 499, L179

\bibitem[{{Mattox} {et~al.}(1996){Mattox}, {Bertsch}, {Chiang}, {Dingus}, {Digel}, {Esposito}, {Fierro}, {Hartman}, {Hunter}, {Kanbach}, {Kniffen}, {Lin}, {Macomb}, {Mayer-Hasselwander}, {Michelson}, {von Montigny}, {Mukherjee}, {Nolan}, {Ramanamurthy}, {Schneid}, {Sreekumar}, {Thompson}, \& {Willis}}]{mattox96}
{Mattox}, J.~R., {Bertsch}, D.~L., {Chiang}, J., {et~al.} 1996, \apj, 461, 396

\bibitem[{{Nishiuchi} {et~al.}(2001){Nishiuchi}, {Yokogawa}, {Koyama}, \& {Hughes}}]{nishiuchi01}
{Nishiuchi}, M., {Yokogawa}, J., {Koyama}, K., \& {Hughes}, J.~P. 2001, \pasj, 53, 99

\bibitem[{{Rakowski} {et~al.}(2003){Rakowski}, {Ghavamian}, \& {Hughes}}]{rakowski03}
{Rakowski}, C.~E., {Ghavamian}, P., \& {Hughes}, J.~P. 2003, \apj, 590, 846

\bibitem[{{Seward} {et~al.}(2012){Seward}, {Charles}, {Foster}, {Dickel}, {Romero}, {Edwards}, {Perry}, \& {Williams}}]{seward12}
{Seward}, F.~D., {Charles}, P.~A., {Foster}, D.~L., {et~al.} 2012, \apj, 759, 123

\bibitem[{{Sreekumar} {et~al.}(1992){Sreekumar}, {Bertsch}, {Dingus}, {Fichtel}, {Hartman}, {Hunter}, {Kanbach}, {Kniffen}, {Lin}, {Mattox}, {Mayer-Hasselwander}, {Michelson}, {von Montigny}, {Nolan}, {Pinkau}, {Schneid}, \& {Thompson}}]{sreekumar92}
{Sreekumar}, P., {Bertsch}, D.~L., {Dingus}, B.~L., {et~al.} 1992, \apjl, 400, L67

\bibitem[{{Toledo-Roy} {et~al.}(2009){Toledo-Roy}, {Vel{\'a}zquez}, {de Colle}, {Gonz{\'a}lez}, {Reynoso}, {Kurtz}, \& {Reyes-Iturbide}}]{toledo09}
{Toledo-Roy}, J.~C., {Vel{\'a}zquez}, P.~F., {de Colle}, F., {et~al.} 2009, \mnras, 395, 351

\bibitem[{{Tramacere} {et~al.}(2016){Tramacere}, {Paraficz}, {Dubath}, {Kneib}, \& {Courbin}}]{Tramacere:2016}
{Tramacere}, A., {Paraficz}, D., {Dubath}, P., {Kneib}, J.~P., \& {Courbin}, F. 2016, \mnras, 463, 2939

\bibitem[{Tramacere \& Vecchio(2013)}]{Tramacere:2013}
Tramacere, A. \& Vecchio, C. 2013, Astronomy and Astrophysics, 549, A138

\bibitem[{{Whelan} {et~al.}(2014){Whelan}, {Kavanagh}, {Sasaki}, {Haberl}, {Maggi}, {Filipovi{\'c}}, {Bozzetto}, \& {Crawford}}]{whelan14}
{Whelan}, E., {Kavanagh}, P., {Sasaki}, M., {et~al.} 2014, in The X-ray Universe 2014, ed. J.-U. {Ness}, 333

\bibitem[{{Williams} \& {Chu}(2005)}]{williamschu05}
{Williams}, R.~M. \& {Chu}, Y.~H. 2005, \apj, 635, 1077

\bibitem[{{Williams} {et~al.}(1997){Williams}, {Chu}, {Dickel}, {Beyer}, {Petre}, {Smith}, \& {Milne}}]{williams97}
{Williams}, R.~M., {Chu}, Y.-H., {Dickel}, J.~R., {et~al.} 1997, \apj, 480, 618

\bibitem[{{Yamaguchi} {et~al.}(2014){Yamaguchi}, {Badenes}, {Petre}, {Nakano}, {Castro}, {Enoto}, {Hiraga}, {Hughes}, {Maeda}, {Nobukawa}, {Safi-Harb}, {Slane}, {Smith}, \& {Uchida}}]{yamag14}
{Yamaguchi}, H., {Badenes}, C., {Petre}, R., {et~al.} 2014, \apjl, 785, L27

\bibitem[{{Yew} {et~al.}(2021){Yew}, {Filipovi{\'c}}, {Stupar}, {Points}, {Sasaki}, {Maggi}, {Haberl}, {Kavanagh}, {Parker}, {Crawford}, {Vukoti{\'c}}, {Uro{\v{s}}evi{\'c}}, {Sano}, {Seitenzahl}, {Rowell}, {Leahy}, {Bozzetto}, {Maitra}, {Leverenz}, {Payne}, {Park}, {Alsaberi}, \& {Pannuti}}]{yew21}
{Yew}, M., {Filipovi{\'c}}, M.~D., {Stupar}, M., {et~al.} 2021, \mnras, 500, 2336

\bibitem[{{Zangrandi} {et~al.}(2024){Zangrandi}, {Jurk}, {Sasaki}, {Knies}, {Filipovi{\'c}}, {Haberl}, {Kavanagh}, {Maitra}, {Maggi}, {Saeedi}, {Bernreuther}, {Koribalski}, {Points}, \& {Staveley-Smith}}]{eros24}
{Zangrandi}, F., {Jurk}, K., {Sasaki}, M., {et~al.} 2024, \aap, 692, A237

\end{thebibliography}

\begin{appendix}
\section{DENCLUE}
\label{denclue-appendix}
The DENCLUE algorithm relies on the localization of local maxima in datasets, using a kernel density estimator. 
Assume that we have a set of instances in a given data space, $D=\{p_1,..,p_n\}$, and  each instance $p_j \in  D$ is a point whose coordinates are represented by the vector  $\mathbf{q}_j$, with dimensionality $d=2$, then the probability density function at a position $p$ in the data space, $f(p)$, can  be expressed in  terms of a kernel function $G$ as
\begin{equation}
f(p)=\frac{1}{nh^2}\sum_{i=1}^{n}  G\big(\frac{\mathbf{q}-\mathbf{q}_i}{h}\big)	
\end{equation}
where $h$ is the kernel bandwidth.
The function $G$ represents the \emph{influence} of the point  $p_i\in D$ with coordinates $\mathbf{q}_i$ at the position $p$  with coordinates $\mathbf{p}$, hence, the probability density function represents the sum of the influence of all the data points on the point  $p$. 
The DENCLUE algorithm  is designed to find for each point $p_ j\in D$ the corresponding   {density attractor} point $p_j^*$, i.e  a local maximum
of $f$. To find the attractors, rather than using a computationally expensive gradient  ascent approach, we use the fast hill climbing technique presented in  int \cite{Hinneburg:2007tq} that is an iterative  update rule with the formula:
\begin{equation}
\mathbf{q}^{(t+1)}=\frac{ \sum_{i=1}^{n} G\big(\frac{\mathbf{q}^{(t)}-\mathbf{q}_i}{h}\big)\mathbf{q}_i }
{ \sum_{i=1}^{n} G\big(\frac{\mathbf{q}^{(t)}-\mathbf{q}_i}{h}\big) }
\end{equation}
where the $t$ is the current iteration, $t+1$ the updated value, and $\mathbf{q}^{(t=0)}\equiv\mathbf{q}_j$ represent the coordinate vector of the point $p_j$.
The fast hill-climbing starts at each point with coordinate vector $\mathbf{q}_j$,  and iterates until $\|\mathbf{q}^{(t)}-\mathbf{q}^{(t+1)} \| \leq \varepsilon_d $. 
The coordinate  vector $\mathbf{q^{(t+1)}}$ identifies  the position of the \emph{density attractor} $p_j^*$ for the point $p_j$. All the attractor points are then clustered using the DBSCAN algorithm, to eventually deblend the confused sources, in a way that can be summarized by the following steps:
\begin{itemize}
\item each cluster $C_{m}$, detected by the DBSCAN, is defined by a set of points $\{p_j \in C_{m}\}$, corresponding to the coordinates of the photons in the cluster 
\item for each point $p_j \in C_{m}$  we compute the  \emph{density attractor}. It means that $\forall p_j \in C_{m}, \exists p_m^*$, i.e. the two sets $\{p_m \}$ and $\{p_m^*\}$, map biunivocally $C_{\rm cls}$ to $C^*_{m}$.
\item all the attractors points in $C^*_{m}$ are clustered using the DBSCAN, in clusters of \emph{density attractors} producing a list of clusters of attractors  $\{C_{A_1},...,C_{A_n}\}$.  
\item The set of points  $p_j$ whose  \emph{density attractors} belong to the same cluster of  attractors $A_{C_n}$, i.e. $\{p_j:p^*_j \in CA_n \}$, defines a new sub-cluster $C_{D_n}$ 
\end{itemize}
\end{appendix}

\end{document}